\documentclass{article}

{}
{}
{}

\usepackage{amsmath} 
\usepackage{amsthm} 
\usepackage{cite} 
\usepackage{hyperref} 
\usepackage{graphicx}
\usepackage{algorithmic} 
\usepackage{hyperref}
\usepackage{authblk}
\usepackage[margin=1.2cm]{geometry}
\usepackage{multicol}
\usepackage{blindtext}
\usepackage[nolist]{acronym}
\usepackage{epstopdf}
\usepackage{float}
\begin{document}





\begin{acronym}
\acro{dss}[DSS]{Decision support System}
\acro{adtree}[ADTree]{Attack-Defense Tree}
\acro{ot}[OT]{Operational Technology}
\acro{it}[IT]{Information Technology}
\acro{et}[ET]{Electrical Technology}
\acro{ics}[ICS]{Industrial Control System}
\acro{ids}[IDS]{Intrusion Detection System}
\acro{pera}[PERA]{Purdue Enterprise Reference Architecture}
\acro{soc}[SOC]{Security Operation Center}
\acro{mcdm}[MCDM]{Multi-Criteria Decision-Making}
\acro{siem}[SIEM]{Security Information and Event Management}
\acro{cti}[CTI]{Cyber Threat Intelligence}
\acro{dst}[DST]{Dempster Shafer Theory}
\acro{ml}[ML]{Machine Learning}
\acro{ioa}[IOA]{Information Object Address}
\acro{c2}[C2]{Command \& Control}
\acro{ioc}[IOC]{Indicator of Compromise}

\end{acronym}

\title{Towards a Comprehensive Framework for Cyber-Incident Response Decision Support in Smart Grids
\thanks{\begin{tiny}This paper is a preprint of a paper submitted to 14th Mediterranean Conference on Power Generation Transmission, Distribution and Energy Conversion (MEDPOWER 2024) and is subject to Institution of Engineering and Technology Copyright. If accepted, the copy of record will be available at IET Digital Library\end{tiny}}
}

\author[1, 2]{Ömer Sen}
\author[2]{Yanico Aust}
\author[2]{Martin Neumüller}
\author[1, 2]{Immanuel Hacker}
\author[1, 2]{Andreas Ulbig}

\affil[1]{{Fraunhofer FIT, Aachen, Germany,}  *Email:  \{oemer.sen, immanuel.hacker\}@fit.fraunhofer.de \newline}
\affil[2]{{RWTH Aachen University, Aachen, Germany,} *Email:\{yanico.aust, martin.neumueller\}@rwth-aachen.de \newline}
\date{}

\maketitle

\begin{abstract}
The modernization of power grid infrastructures necessitates the incorporation of decision support systems to effectively mitigate cybersecurity threats. This paper presents a comprehensive framework based on integrating Attack-Defense Trees and the Multi-Criteria Decision Making method to enhance smart grid cybersecurity. By analyzing risk attributes and optimizing defense strategies, this framework enables grid operators to prioritize critical security measures. Additionally, this paper incorporates findings on decision-making processes in intelligent power systems to present a comprehensive approach to grid cybersecurity. The proposed model aims to optimize the effectiveness and efficiency of grid cybersecurity efforts while offering insights into future grid management challenges.
\end{abstract}

\begin{multicols}{2}

\section*{Introduction} \label{sec:introduction}
The rapid evolution of smart grid technology has brought about a transformative shift in how traditional power distribution systems operate. The integration of renewable energy sources and information technology has increased the grid's complexity and its susceptibility to cyberattacks. Incidents like the 2015 Ukraine power grid attack, which caused significant power outages affecting hundreds of thousands of customers, underscore the urgency of establishing robust cybersecurity frameworks. Graph-based playbooks provide a comprehensive graphical tool for visualizing potential threats and defenses, enhancing a system's capacity to respond effectively to incidents.

Smart grids, by nature of their digital evolution, have transformed traditional power distribution, improving efficiency in energy management and distribution. However, this digital evolution has also introduced increased vulnerabilities, particularly concerning cybersecurity. Modern power grids face an evolving landscape of cyber threats, where traditional incident response methods have proven inadequate in addressing these challenges. \acp{dss} are essential for bridging the gaps in incident response, aiding security operations in identifying and mitigating potential threats. These systems enable operators to visualize, analyze, and respond to complex threat scenarios in real-time, offering a proactive approach to securing the grid infrastructure.

In particular, the challenges that arise with a comprehensive incident response decision guide in critical situations involve the perception of the incident in a form that is comprehensively detailed to augment the correct countermeasures. Situational awareness for incident recognition is the first step that enables the application of accurate and organization-specific countermeasure strategy identification. Furthermore, different criteria for decision-making need to be consolidated to identify the optimal set of countermeasures addressing the detected incident, which, in a time-critical situation, also needs to be automated to support critical decision-making within short time periods.

The primary objective of this paper is to present a comprehensive framework for \ac{dss}s specifically designed for smart grid environments. The framework integrates advanced methodologies like \acp{adtree} and \acp{mcdm} method to assess and mitigate risks. This approach provides smart grid operators with a systematic method to identify optimal defense strategies and address vulnerabilities effectively. Moreover, the framework emphasizes the need for predictive analytics to preemptively identify emerging threats, allowing for a proactive defense mechanism that can adapt to the evolving threat landscape.

\vspace{-1em}
\section*{Background} \label{sec:background}
\vspace{-1em}
\subsection*{Multi-Stage Cyberattacks} \label{subsec:background_attack}
Most of the cyberattacks carried out in recent years do not follow a simple attack structure but rather consist of a combination of smaller steps. A composition of these smaller steps is usually referred to as a cyber kill-chain. Although this concept originated from the U.S. military \cite{bryant2017novel}, the classical term for a cyber kill-chain is often derived from the Lockheed Martin kill-chain \cite{hutchins2011intelligence}. It describes a sequence of seven distinct steps, which serve as a descriptive guide for a successful attack. These steps typically progress from the attacker's weak positioning to the most impactful position. The focus on multi-stage cyberattacks has many fields of application, including the research of existing security frameworks, the categorization of existing attacks, understanding future and yet-to-exist attack vectors, and the interplay between different network layers.

A well-researched approach is provided by the MITRE ATT\&CK ICS framework, which covers 12 tactics encompassing a total of 83 techniques. Compared to previous attempts, the MITRE ATT\&CK ICS framework outlines possible movements between the \ac{it}, \ac{ot}, and \ac{et} layers of an industrial network. This approach aligns with the nature of modern cyberattacks, in which the attack structure is composed of multiple levels, layers, or privileges that must be obtained to achieve the final target, hence the term multi-stage cyberattacks. Designing such attack vectors often involves the Purdue model \cite{assante2015industrial}. The different stages of such attacks are then tailored concerning the devices and applications relying on \ac{it} and \ac{ot} zones in \acp{ics}.

An exemplary depiction of the well-known Stuxnet multi-stage cyberattack was analyzed and provided by MITRE. They provided the kill-chain for enterprise networks as well as for \ac{ics} networks\footnote{MITRE's kill-chain for the Stuxnet attack: \url{https://attack.mitre.org/software/S0603/} [online: 07.08.2024]}.
\vspace{-1em} 
\subsection*{Purdue Model} \label{subsec:background_purdue}
The Purdue model is used to depict the architecture of an \ac{ics} network and to categorize industrial equipment, operations, networks, and functions into a hierarchical structure \cite{assante2015industrial, williams1994purdue, williams1996overview}. It classifies the \ac{ics} network architecture into \ac{ot} and \ac{it} zones, which are further divided into six levels, ranging from physical hardware to enterprise network integration \cite{williams1994purdue, williams1996overview}. The goal of the Purdue model is to separate the networks within an \ac{ics} into industrial and traditional networks, providing an extended layering and clear distinction between the devices used and the network access possibilities.

\vspace{-1em}
\subsection*{Related Work} \label{subsec:background_relatedwork}
The model components required to provide cyber-incident responses can be broken down into the following three: an \ac{ids}, which creates security events; a \ac{siem} system, which processes these events and correlates them accordingly; and finally a \ac{dss}, which helps identify and mitigate potential threats. The selection of the first component depends on the chosen system and is therefore use-case specific. The implementation by Sen et al. \cite{sen2022using} detects the corresponding attack evolution and strategy based on domain-specific attribution and contextual correlation of cyber incident indicators. We will follow and extend their implementation by enabling the interplay between a low-level \ac{ids}, forwarding the alerts to the \ac{siem}, and allowing for high-level correlation \cite{sen2022using}.

In the context of \ac{siem} systems, several attempts have been made to assist in providing process awareness. Bryant et al. \cite{bryant2017novel} compared their Bryant kill-chain approach to the Lockheed Martin kill-chain, most commonly known as a classical cybersecurity kill-chain. Using their approach, they address one of the common problems of detail enrichment for alarms. Although novel, their work is not aimed at application in the industrial context, which is a limitation for our investigation. Additionally, our current approach is implemented to work under MITRE's ATT\&CK \ac{ics} kill-chain, which is not considered by Bryant et al. \cite{bryant2017novel}. Motlhabi et al. \cite{motlhabi2022context} described the idea of a \ac{cti} exchange platform to assist with further analysis by \ac{soc} personnel. Our work provides an exemplary implementation of their structure, with additional features such as sophisticated attack correlation to attack actions, kill-chain identification based on the MITRE ATT\&CK \ac{ics} framework, and the ability to reconstruct the attack.

A survey of different \ac{siem} implementations aimed at providing situational awareness has been conducted by Ünal et al. \cite{unal2021investigation}. Many publications focus on accumulating information before or in parallel with the \ac{siem} system, such as \ac{ml}-based anomaly detection \cite{hindy2018improving, moukafih2020neural, vasilyev2020security}. Regarding the successive step to \ac{siem}'s data correlation, the landscape they covered only included a \ac{soc}-based approach \cite{hwoij2021siem}, which opens the door for our \ac{dss}-based approach.

When considering a \ac{dss}, the two main differences highlighted in the literature are the type of attack modeling and the countermeasure provision techniques used. Most approaches utilize attack graphs for attack modeling \cite{kotenko2016dynamical, shameli2016dynamic}. An attack graph represents all possible steps and paths an attacker can take to reach a certain goal. An attack graph can also be extended to a Bayesian attack graph, which includes the probabilities of transitions between steps, as implemented in \cite{miehling2015optimal}. Depending on the implementation of the attack graph, the nodes can also include further information, such as kill chain steps or techniques according to the MITRE ATT\&CK framework.

To better assess the impact of countermeasures, some approaches include service dependency graphs \cite{kotenko2016dynamical, shameli2016dynamic}. Service dependency graphs show the dependencies between services, allowing for the assessment of the impact and cost of a countermeasure on the services. The countermeasure provision techniques differ in how they select the countermeasures. Heuristic optimization methods are used in \cite{kotenko2016dynamical}, allowing for the efficient search of large solution spaces, but they can become resource-intensive for larger systems. In another approach \cite{shameli2016dynamic}, simple additive weighting methods are used to better select countermeasures to suit the needs of the environment by prioritizing the countermeasures based on security benefits, security impact, and security cost. Following that, based on the Pareto Optimal Set Selection method, the countermeasures are selected to provide the best security benefits with the least security impact and cost.

In contrast, our approach utilizes a Bayesian attack graph that includes attack actions, kill chain steps, and the origin and target of the attack. Based on a simple additive weighting method and an \ac{mcdm} method, the countermeasures are selected to provide the best security benefits for the environment.
\vspace{-1em} 
\section*{Methodology} \label{sec:methodology}
\vspace{-1em}
\subsection*{Overview} \label{subsec:methdology_overview}
The methodology integrates advanced tools and standards to enhance cybersecurity in smart grids. It uses \acp{adtree} to graphically represent threats and defense strategies, helping to break down complex attack objectives into manageable components. The \ac{mcdm} method is employed to understand the influence of risk attributes on overall system risk. This analysis identifies critical factors, guiding the prioritization of cybersecurity resources.

The framework adopts structured threat modeling to identify, assess, and prioritize potential vulnerabilities in smart grids. This approach emphasizes the importance of understanding data flow and system interactions to accurately assess potential risks. By focusing on risk attributes such as probability, impact, and cost, the framework ensures that cybersecurity measures are both efficient and effective.
\vspace{-0.75em}
\begin{figure}[H]
	\centerline{\includegraphics[width=\linewidth]{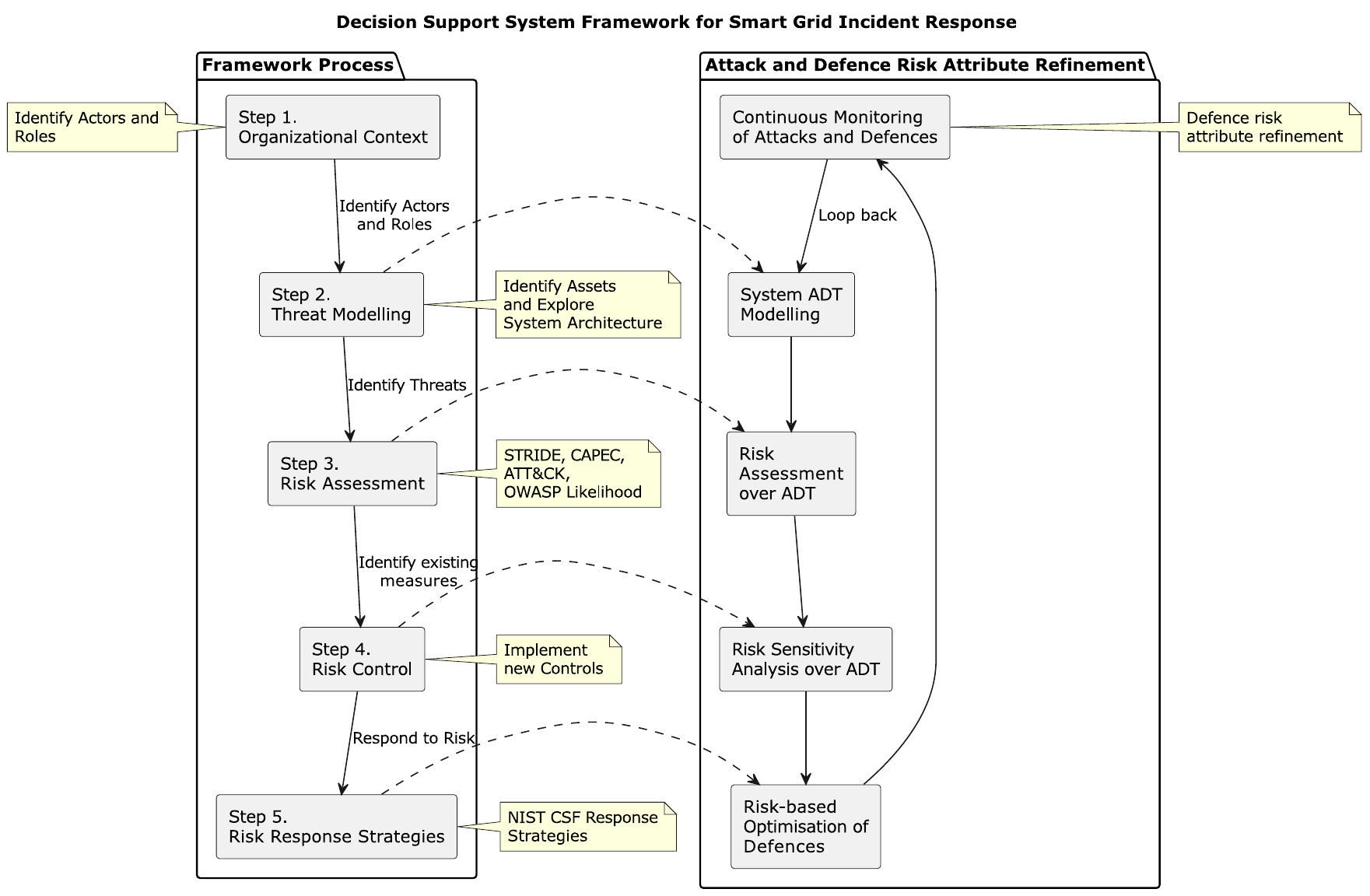}}
	\caption{
		\ac{dss} Framework for Smart Grid Incident Response. This diagram outlines the key processes involved in the \ac{dss} framework for incident response in smart grids. The framework consists of five main steps: organizational context, threat modeling, risk assessment, risk control, and risk response strategies. Additionally, the attack and defense risk attribute refinement process involves system \ac{adtree} modeling, risk assessment over \ac{adtree}, risk sensitivity analysis, risk-based optimization of defenses, and continuous monitoring. This holistic approach aims to enhance the cybersecurity posture of smart grids by identifying, evaluating, and mitigating potential risks effectively.
	}
	\label{fig:framework}
\end{figure}
\vspace{-1em}
Risk management involves a systematic approach to determine the likelihood and impact of identified threats. This includes analyzing various factors that contribute to the probability of an attack, as well as evaluating the potential business and technical impacts. The resulting risk assessment helps guide the selection of appropriate cybersecurity measures. A comprehensive risk management strategy includes several key components (cf. Figure~\ref{fig:framework}).






\vspace{-1em}
\subsection*{Multi-Staged Attack Detection} \label{subsec:methodology_attack}
\begin{figure}[H]
    \centerline{\includegraphics[width=1\linewidth]{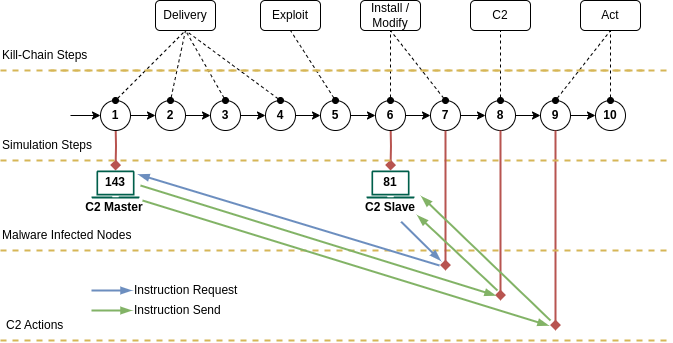}}
    \caption{Visualization of a Havex-based attack, following the classical Lockheed Martin kill-chain. The simulation concludes in 10 steps, with each step corresponding to a single kill-chain stage. The attack is based on two \ac{c2} nodes, the first being the master and the second the slave. As indicated in simulation steps 7, 8, and 9, communication attempts are initiated.}
    \label{fig:havex_killchain}
\end{figure}

To better test our \ac{dss}, we designed and performed attacks using Stuxnet and Havex techniques. We depicted the Havex attack in Figure \ref{fig:havex_killchain}. Our goals were to 1) provide an attack according to the Lockheed Martin kill-chain, 2) ensure the attack progresses through these steps sequentially, 3) include infected nodes, and 4) target specific nodes.


The attack we designed performs 10 simulation steps, from beginning to completion. Figure \ref{fig:havex_killchain} depicts the simulation steps and their correlation to the kill-chain, depending on the action taken. The attack is structured to deploy a \ac{c2} master in the first simulation step, which delegates further commands. In step 6, a node becomes infected as a \ac{c2} slave. The following steps (7, 8, 9) are designed to request information from the \ac{c2} master and receive instructions. Our approach involves a locally developed \ac{ids}, which monitors traffic at the lower sensory level, triggering on \ac{it}-based and \ac{ot}-based events such as changes in IP or MAC addresses, mismatched ports, incorrect connections, changes in expected \acp{ioa}, or even mismatches in sequence controls related to the I-frame or S-frame of the IEC 104 protocol. These \acp{ioc}, along with the created events, are forwarded to our \ac{siem} system, which investigates the incoming events according to known attack patterns.
\vspace{-0.75em}

The \ac{siem} processes these events and correlates them with known attack graphs. Once the collected \acp{ioc} strongly correlate with a known attack graph, the attack graph is forwarded to the \ac{dss}, which aids in identifying and mitigating potential threats. In our example involving the Havex attack, the \ac{siem} correlates the attack graph depicted in Figure \ref{fig:attack_graph}.

\vspace{-1em}
\subsection*{Multi-Criteria Decision Support} 
\begin{figure}[H]
    \centerline{\includegraphics[width=1\linewidth]{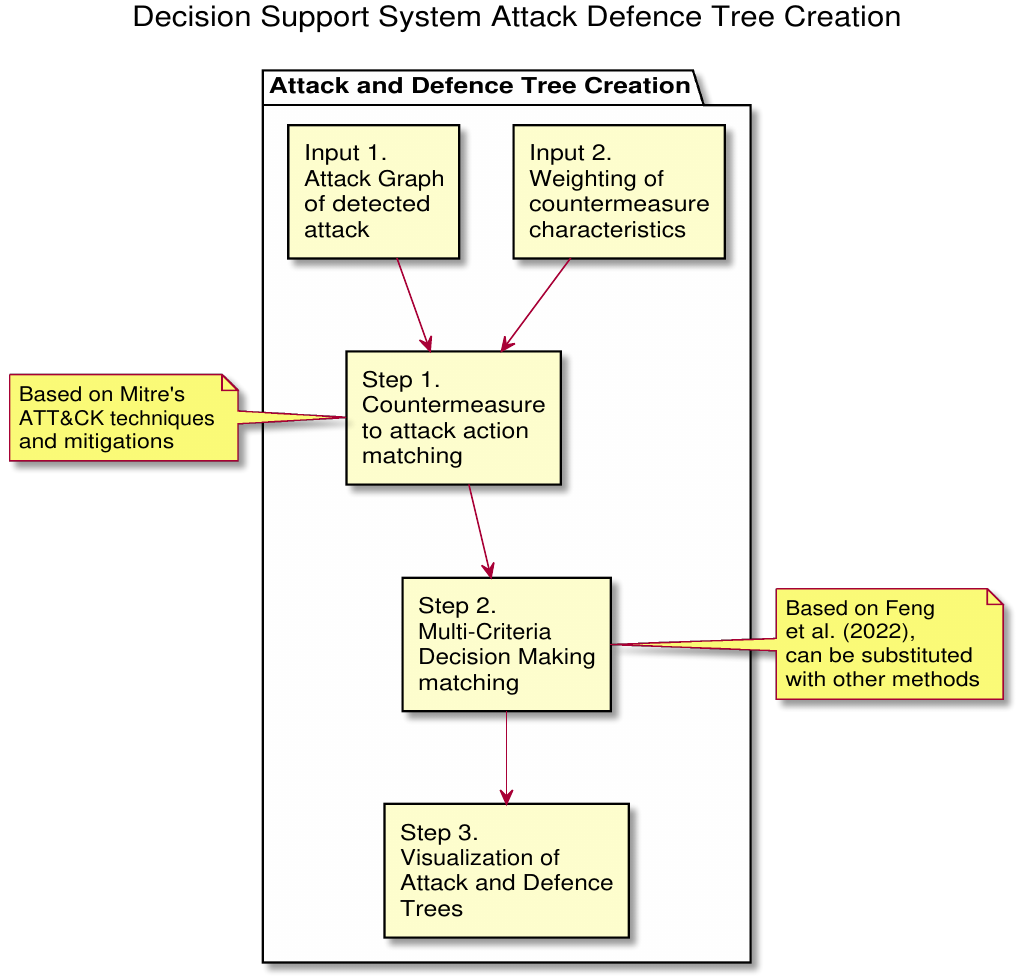}}
    \caption{Overview of the Multi-Criteria \ac{dss}. This diagram illustrates the key components of the \ac{dss} for incident response in smart grids. The system consists of three main components: Countermeasure to attack action matching, Multi-Criteria Decision Making, and Visualization of the \acp{adtree}. This approach aims to provide grid operators with a decision-making tool to effectively respond to cybersecurity incidents.}
    \label{fig:decision_support}
\end{figure}

\vspace{-1em}
\label{subsec:methodology_DSS}
The goal of the \ac{dss} is to provide grid operators with countermeasures to effectively respond to cybersecurity incidents based on multiple criteria. The \ac{dss} consists of three key components, as seen in Figure~\ref{fig:decision_support}: Countermeasure to attack action matching, Multi-Criteria Decision Making, and Visualization of the \acp{adtree}.

\begin{figure}[H]
    \centerline{\includegraphics[clip, trim=0 7cm 0 0, width=1\linewidth]{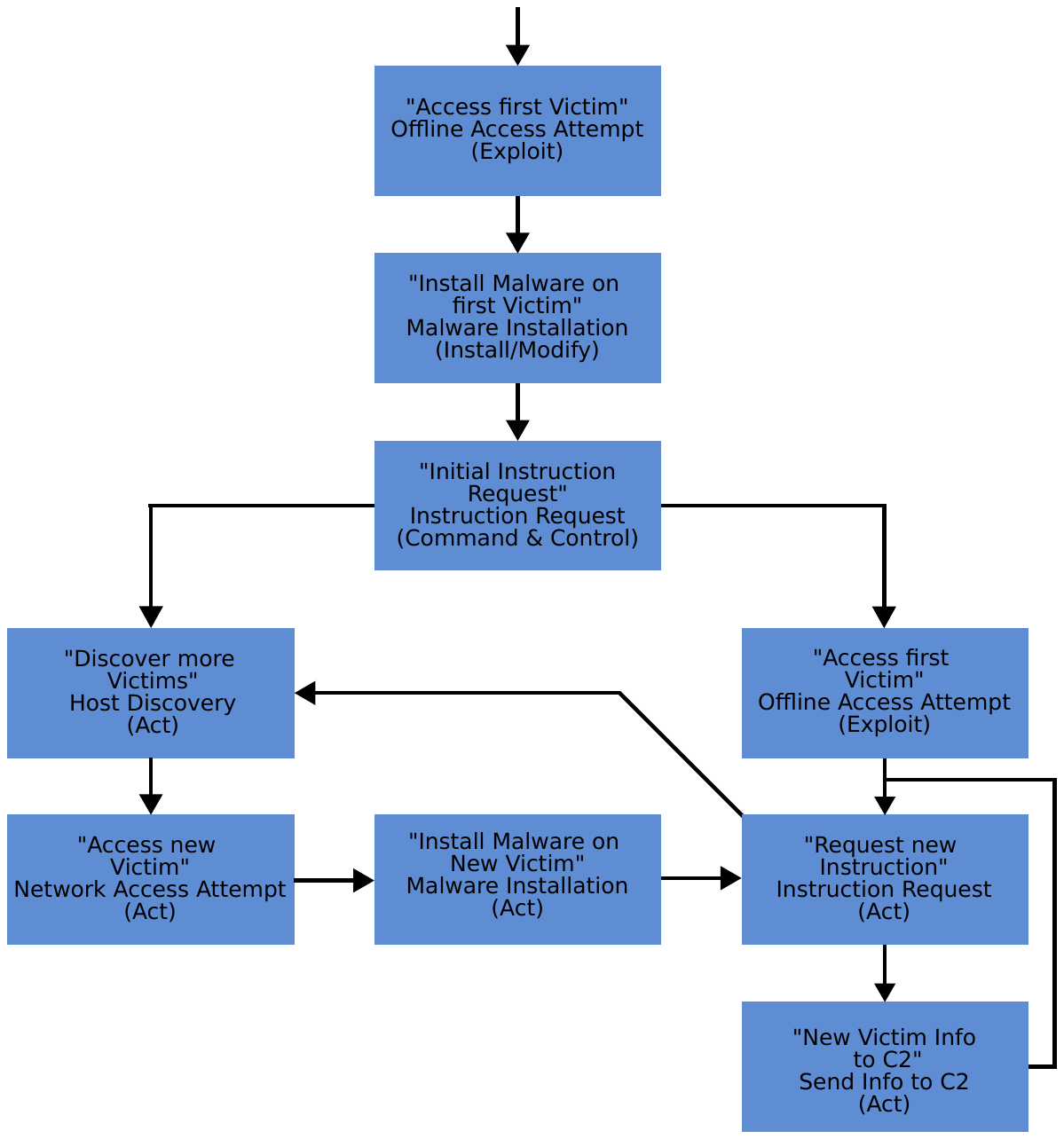}}
    \caption{Attack Graph of the Havex Malware. Each node includes the name of the step, the attack action, and its kill-chain step. The edges include the mass function of the connection and the properties of the origin and target. The graph is used as input for the \ac{dss}.}
    \label{fig:attack_graph}
\end{figure}

Firstly, as input to the system, the weightings of the criteria are provided. These criteria include the cost to implement, time to implement, local or remote setup, duration of activation, area of impact, and technical impact. Depending on the environment, the grid operator can prioritize these criteria by weighting them according to their importance, leading to a more effective response to the incident. Secondly, the attack graph detected in the previous Multi-Staged Attack Detection step is given as input. An example attack graph of the Havex attack, which can be seen in Figure~\ref{fig:attack_graph}, contains the attack actions, the kill-chain step, the mass function of the connection, and the properties of the origin and target. With this information, the system performs three steps.

\textbf{Step 1. Countermeasure to attack action matching}. Based on the MITRE ATT\&CK framework, the attack actions are matched to the countermeasures. For this, we created a list that includes all techniques in the ATT\&CK framework related to \acp{ics}. For each technique, ATT\&CK provides countermeasures that we add to each technique in our list. Additionally, each countermeasure has values for each of the criteria. Based on the techniques, the corresponding countermeasures are extracted to be used in the next step.
\vspace{-0.75em}

\textbf{Step 2. Multi-Criteria Decision Making}. In this step, the countermeasures are evaluated based on the criteria. Based on the weightings given as input and the values of the criteria for each countermeasure, Multi-Criteria Decision Making is performed to rank the countermeasures. This can be done by a variety of methods and is easily interchangeable in our system. For now, we use an Interval-valued Pythagorean fuzzy multi-criteria decision-making method based on the set pair analysis theory and Choquet integral \cite{li2023interval}. This creates a ranking of the countermeasures based on the criteria and their weightings. Based on this ranking, we choose the countermeasure with the highest score for each attack action to be used in the next step.

\textbf{Step 3. Visualization of the \acp{adtree}}. Using the best-ranked countermeasures for each attack action, we create an \ac{adtree} based on the attack graph given as input, enhanced with the countermeasures. This \ac{adtree} is then visualized for the grid operator to provide an overview of the attack and the countermeasures that can be taken. This visualization can then be used to make an informed decision in a timely manner to respond to the incident.

\vspace{-0.75em}

\vspace{-3em}
\section*{Result} \label{sec:result}
\vspace{-1em}
\subsection*{Scope} \label{subsec:result_proc}
The expected outcome of this research is a structured decision support framework that enhances the cybersecurity posture of smart grid systems. By integrating principles of intelligent decision-making into \ac{adtree} and applying the \ac{mcdm} method, the framework provides a comprehensive approach to assessing and mitigating risks. It includes detailed scenarios of multi-stage cyberattacks to improve situational awareness and detection. The framework augments \ac{adtree} to systematically identify and evaluate potential countermeasures. Through the application of \ac{mcdm}, it prioritizes the best set of countermeasures based on defined metrics such as risk severity and cost-effectiveness. A sensitivity analysis further examines how different weightings of criteria impact decision-making, ensuring the robustness of the chosen strategies.

Additionally, the framework presents a graph-based representation of playbooks to visually guide operators through recommended response actions. Designed to be dynamic and continuously evolving, the framework adapts to new cyber threats, enabling smart grid operators to make data-driven decisions that prioritize security measures effectively.

Ultimately, this research aims to contribute to the development of more resilient smart grid systems capable of withstanding modern cyber threats. By emphasizing continuous improvement in cybersecurity strategies, the framework ensures long-term security and resilience for smart grid infrastructure.

\vspace{-0.75em}
\begin{figure}[H]
    \centerline{\includegraphics[clip, trim= 2.5cm 3cm 2.5cm 4cm, width=1\linewidth]{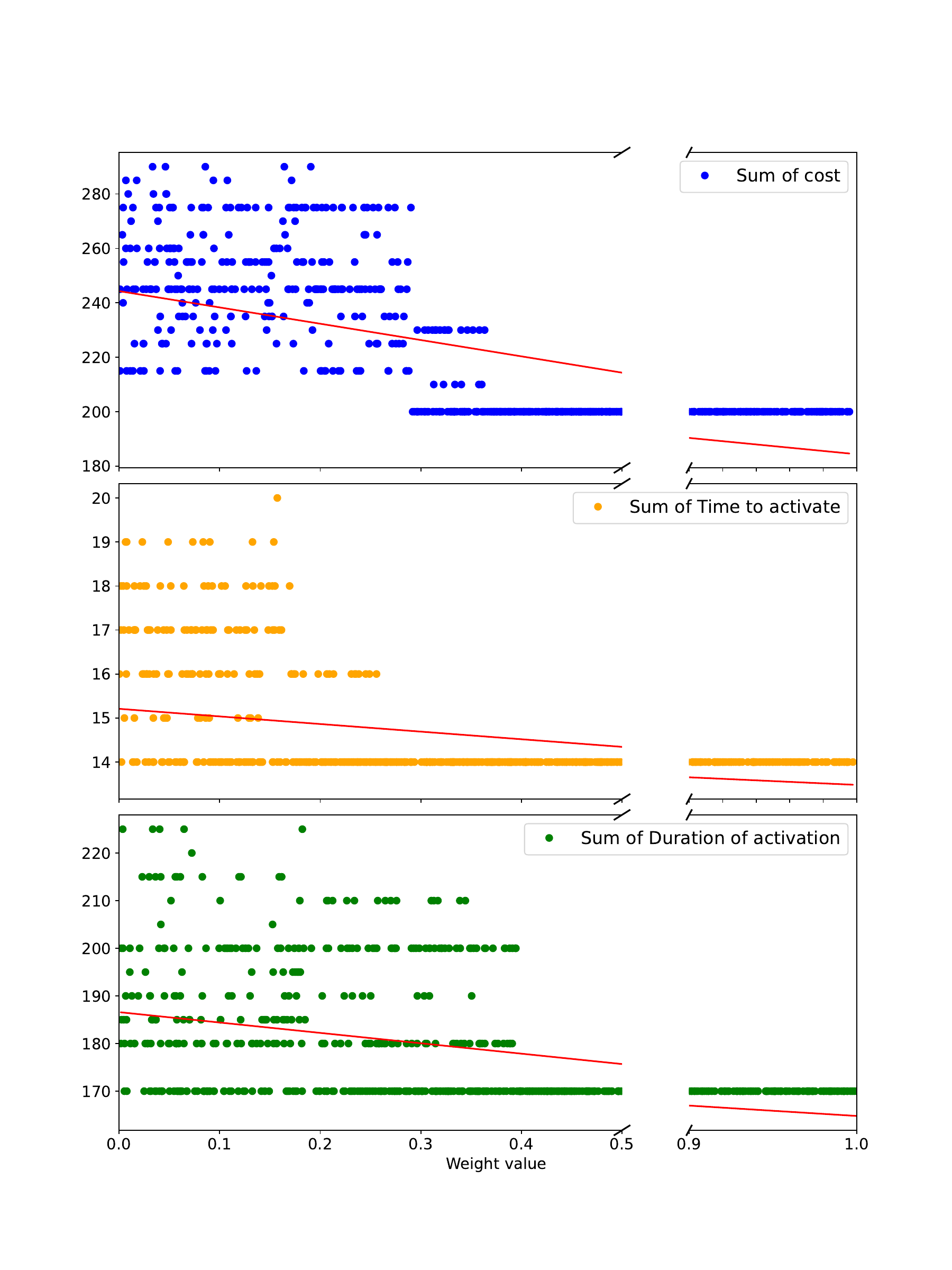}}
    \caption{Each graph shows the correlation between the weight of a criterion and the resulting score. The top graph represents the cost criterion, the middle graph the time to activate criterion, and the bottom graph the duration of activation criterion. The x-axis shows the value of the weight of the criterion, and the y-axis shows the sum of the scores of the countermeasures. The red line shows the trend line of the correlation between the weight of the criterion and the resulting score. On the x-axis, between 0.5 and 0.9, the sums remain constant; for better readability, we have truncated that part.}
    \label{fig:results_weight_to_score}
\end{figure}

\vspace{-2.5em}
\subsection*{Outcome} \label{subsec:result_outcome}
\vspace{-1em}
Due to the varying needs of each environment, it is challenging to provide a general recommendation for countermeasures. As a result, the quality of the resulting countermeasures cannot be broadly evaluated. However, the correlation between the weights assigned to the criteria for the countermeasures and the resulting countermeasures can be analyzed. In this research, we examined the impact of the weights of three specific criteria: cost, time to activate, and duration of activation. These criteria were chosen due to the ease of visualizing the results. For these selected criteria, a lower score indicates a better outcome, as the cost should be minimized, the time to activate should be as short as possible, and the duration of activation should also be minimized. The research aimed to demonstrate that the weights of the criteria have a clear correlation with the resulting countermeasures.

For each criterion, we simulated 1000 scenarios with randomly selected weights for the focused criteria and equally distributed weights for the remaining criteria. This approach was chosen to better visualize the results, as visualizing the correlations between all criteria values would result in unclear graphs. The results can be seen in Figure \ref{fig:results_weight_to_score}. Each graph shows the correlation between the weight of a criterion and the resulting score. The top graph represents the cost criterion, the middle graph the time to activate criterion, and the bottom graph the duration of activation criterion. In each graph, the trend line is shown in red, indicating the correlation between the weight of the criteria and the resulting score. 

\vspace{-1.25em}
\begin{figure}[H]
    \centerline{\includegraphics[clip, trim=0 7cm 0 0, width=1\linewidth]{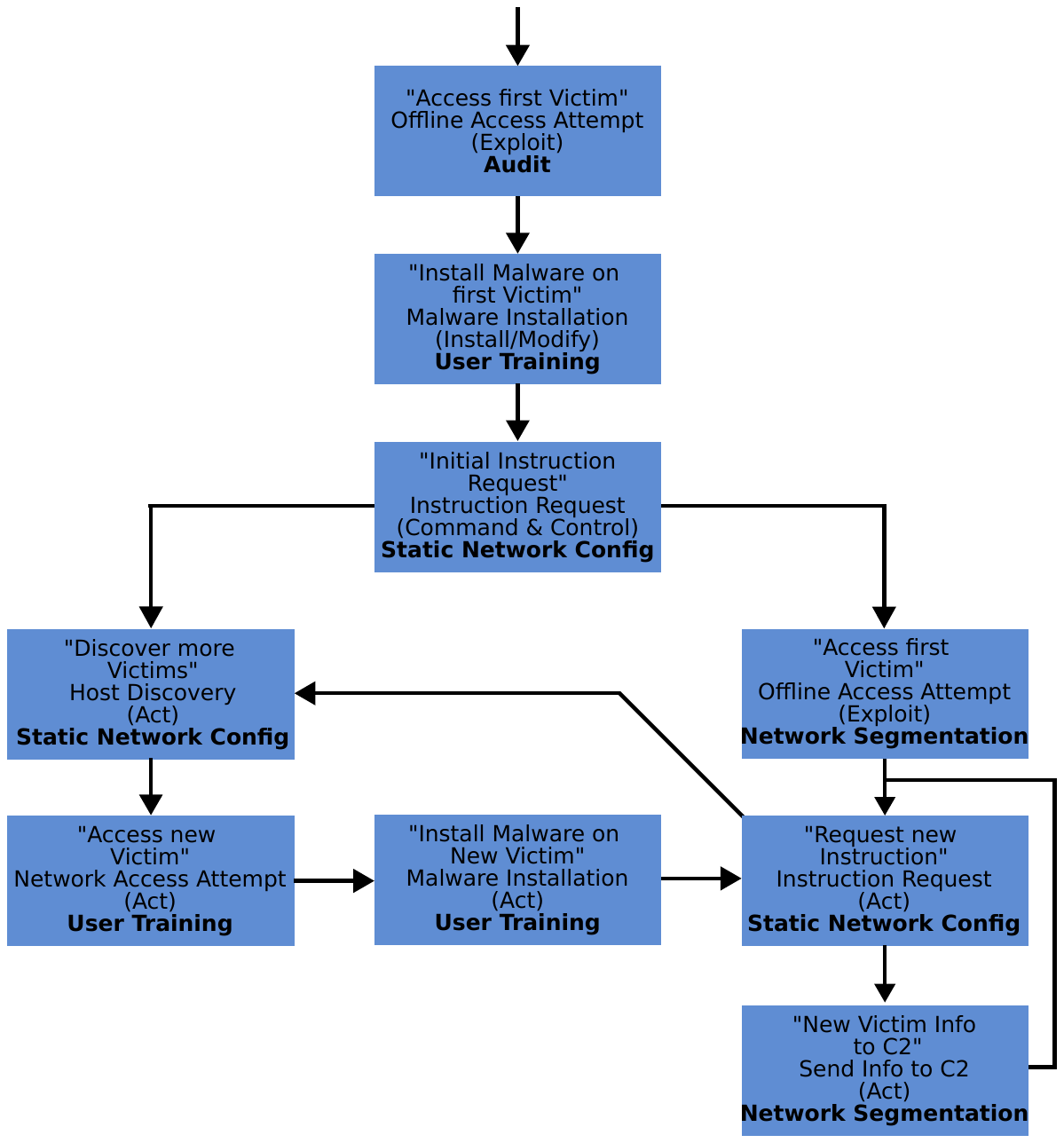}}
    \caption{Example of the resulting countermeasures for the Havex attack. The countermeasures are optimized by the weights of the criteria. The countermeasure for each step is shown at the bottom of each node in bold.}
    \label{fig:results_example}
\end{figure}
\vspace{-1.5em}

It can be clearly seen that the weights of the criteria have a significant negative correlation with the resulting score. This indicates that tuning the weights has a direct impact on the effectiveness of the countermeasures. The resulting countermeasures in all simulations correspond to the possible countermeasures for the Havex attack as described in the MITRE ATT\&CK framework. An exemplary result that an operator would observe is shown in Figure \ref{fig:results_example}. For each step in the attack graph, a corresponding countermeasure optimized by the weights of the criteria is shown. The countermeasure for each step is displayed at the bottom of each node in bold.

\vspace{-1em}
\subsection*{Discussion} \label{subsec:result_discussion}
The results of the research indicate that the weights of the criteria have a clear correlation with the scores of the resulting countermeasures. This finding is significant as it demonstrates that tuning the weights directly impacts the effectiveness of the countermeasures. For grid operators, this is crucial, as it allows them to prioritize countermeasures based on their specific needs and requirements. This approach enables operators to make data-driven decisions, ensuring that their cybersecurity strategies are tailored to the unique demands of their environment.

Interestingly, the optimum for the criteria is often reached at a weighting of around 0.4. A possible explanation for this could be that it reflects the point at which the weighting of the observed criteria surpasses that of the other criteria, influencing the selection of countermeasures. In experiments where all criteria were chosen randomly, the results were less clear compared to those where the weights were assigned in a more structured manner. In the latter case, the optimum was typically reached at a later stage, still showing a clear correlation between the weights of the criteria and the resulting countermeasures. This can be attributed to the nature of multi-criteria decision-making, where the impact of individual criteria on the overall result is complex to visualize.

The results demonstrate that the proposed framework can provide grid operators with an effective decision-making tool for responding to cybersecurity incidents. It ensures that their cybersecurity strategies are customized to meet their individual needs.

\vspace{-1.25em}
\section*{Conclusion} \label{sec:conclusion}
\vspace{-1em}

This paper presents a comprehensive framework to enhance smart grid cybersecurity by integrating situational awareness for incident recognition and a \ac{mcdm} based incident response playbook. The framework enables grid operators to identify and prioritize critical security measures, optimizing defense strategies based on organization-specific attributes. Its modular design allows for the implementation of different \ac{mcdm} methods and supports the inclusion of various sources for techniques-mitigation relations beyond the MITRE framework. This adaptability makes it a practical tool for real-time incident response and ongoing research.

The framework is useful for decision-making, as it can be tailored to the specific needs of different environments. Moreover, its modularity facilitates the easy implementation of different criteria, attack-defense frameworks, and \ac{mcdm} algorithms, making it an invaluable resource for both operational and research purposes. Ultimately, this research contributes to more secure and adaptable smart grid systems, equipping operators with the tools needed for effective cybersecurity management.

\vspace{-1em}
\section*{Acknowledgment}
\begin{minipage}{0.65\columnwidth}%
Received funding from the BMBF under project no. 03SF0694A (Beautiful).
\end{minipage}
\hspace{0.02\columnwidth}
\begin{minipage}{0.23\columnwidth}%
	\includegraphics[width=\textwidth]{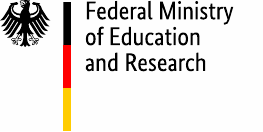}
\end{minipage}
\vspace{-0.5em}

\vspace{-0.5em}
\bibliographystyle{IEEEtran}
\bibliography{conference_101719}
\end{multicols}

\end{document}